\documentclass[runningheads]{llncs}

\usepackage{amsfonts,proof,qtree,amsmath,frameit,mathtools,algpseudocode,algorithm}
\usepackage{latexsym}
\usepackage{graphicx}
\usepackage[usenames,dvipsnames]{color}
\usepackage{listings}
\usepackage{float}
\usepackage{multirow}
\usepackage[scaled]{helvet}

\usepackage{mathrsfs}
\usepackage{mathpartir}
\usepackage{dsfont}
\usepackage{stmaryrd}
\usepackage{url}
\usepackage{textcomp}
\usepackage[colorlinks=true,allcolors=blue,breaklinks,draft=false]{hyperref}
\usepackage{alltt}
\usepackage{bbm}
\usepackage{alltt}
\usepackage{verbdef}
\usepackage{xspace}
\usepackage{verbatim}
\usepackage{enumitem}
\usepackage{lipsum}
\usepackage{wrapfig}
\usepackage[usenames,dvipsnames]{xcolor}
\hypersetup{linkcolor=black,citecolor=black,urlcolor=RubineRed}

\usepackage{enumitem}
\usepackage{cite}
\usepackage{comment}

\let\llncssubparagraph\subparagraph
\let\subparagraph\paragraph
\usepackage{titlesec}
\let\subparagraph\llncssubparagraph

\titlespacing*{\section}{0pt}{*1}{*1}
\titlespacing*{\subsection}{0pt}{*0.7}{*0.5}
\titlespacing*{\paragraph}{0pt}{*0.5}{*0.5}
\renewcommand{\figurename}{Fig.}

\definecolor{shadecolor}{gray}{1.00}
\definecolor{ddarkgray}{gray}{0.75}
\definecolor{darkgray}{gray}{0.30}
\definecolor{light-gray}{gray}{0.87}

\newcommand{\etc}{\emph{etc}}
\newcommand{\ie}{\emph{i.e.}\xspace}

\newcommand{\eg}{\emph{e.g.}\xspace}

\newcommand{\etal}{\emph{et~al.}\xspace}

\newcommand{\fstar}{$\text{F}^{\star}$\xspace}
\newcommand{\dao}{\textsf{DAO}\xspace}
\newcommand{\oraclize}{\textsf{Oraclize}\xspace}
\newcommand{\blockking}{\textsf{BlockKing}\xspace}

\definecolor{pblue}{rgb}{0.13,0.13,1}
\definecolor{pgreen}{rgb}{0,0.5,0}
\definecolor{pred}{rgb}{0.9,0,0}
\definecolor{pgrey}{rgb}{0.46,0.45,0.48}

\definecolor{ckeyword}{HTML}{7F0055}
\definecolor{ccomment}{HTML}{3F7F5F}
\definecolor{cnumber}{HTML}{2A0099}

\lstdefinelanguage{Java}{
  keywords={new, return, public, private, synchronized, void, final, class, new},    
  ndkeywords={bool, int},
  showspaces=false,
  showtabs=false,
  breaklines=true,
  showstringspaces=false,
  breakatwhitespace=true,
  lineskip=-0.6pt,
  basewidth={0.54em, 0.4em},%
  basicstyle=\footnotesize\ttfamily,
  keywordstyle={\color{ckeyword}\ttfamily\bfseries},
  ndkeywordstyle={\color{pblue}\ttfamily\bfseries},
  commentstyle={\color{ccomment}\itshape},
  stringstyle=\color{red},
  moredelim=[il][\textcolor{pgrey}]{$$},
  moredelim=[is][\textcolor{pgrey}]{\%\%}{\%\%}
}

\lstdefinelanguage{Solidity}{
  keywords={typeof, modifier, function, public, returns, external, contract, 
new, true, false, private, catch, function, return, null, throw, catch, switch, var, if, in, while, do, else, case, break},
  ndkeywords={bool, address, mapping, uint, bytes32, string},
  identifierstyle=\color{black},
  sensitive=false,
  comment=[l]{//},
  morecomment=[s]{/*}{*/},
  commentstyle=\color{ccomment}\ttfamily,
  string=[b]",
  showstringspaces=false,
  morestring=[b]',
  showspaces=false,
  showtabs=false,
  breaklines=true,
  morekeywords={function, contract, returns, return},
  breakatwhitespace=true,
  lineskip=-0.6pt,
  basewidth={0.54em, 0.4em},%
  basicstyle=\footnotesize\ttfamily,
  keywordstyle={\color{ckeyword}\ttfamily\bfseries},
  ndkeywordstyle={\color{pblue}\ttfamily\bfseries},
  commentstyle={\color{ccomment}\itshape},
  stringstyle=\color{pred}\ttfamily,
  numberstyle=\scriptsize\color{cnumber}\sffamily,
  moredelim=[il][\textcolor{pgrey}]{$$},
  moredelim=[is][\textcolor{pgrey}]{\%\%}{\%\%}
}

\newcommand{\jcode}[1]{\lstinline[language=Java]{#1}}
\newcommand{\scode}[1]{\lstinline[language=Solidity]{#1}}

\begin{document}

\title{A Concurrent Perspective on Smart Contracts}

% \author{Authors omitted for double-blind review} \institute{}

\author{
  Ilya Sergey\inst{1} \and
  Aquinas Hobor\inst{2}
 %I suppose we might want to consider how to do the author order but I think this idea is mostly yours anyway...
}

\institute{
  University College London, United Kingdom\\
  \email{i.sergey@ucl.ac.uk} \and
  Yale-NUS College and School of Computing, National University of
  Singapore\\
\email{hobor@comp.nus.edu.sg}}

\maketitle

\thispagestyle{plain}
\pagestyle{plain}

\begin{abstract}

In this paper, we explore remarkable similarities between
multi-transactional behaviors of smart contracts in cryptocurrencies
such as Ethereum and classical problems of shared-memory concurrency.
We examine two real-world examples from the
Ethereum blockchain and analyzing how they are vulnerable to
bugs that are closely reminiscent to those that often occur in traditional
concurrent programs.
We then elaborate on the relation between observable contract
behaviors and well-studied concurrency topics, such as
{atomicity}, {interference}, {synchronization}, and
{resource ownership}.
The described \emph{contracts-as-concurrent-objects} analogy provides
deeper understanding of potential threats for smart contracts, indicate
better engineering practices, and enable applications of existing
state-of-the-art formal verification techniques.

\end{abstract}

\section{Introduction}
\label{sec:intro}

Smart contracts are programs that are stored on a blockchain, a distributed
Byzantine-fault-tolerant database.  Smart contracts can be
triggered by blockchain transactions and read and write data on their
blockchain~\cite{ethereum-yellow-paper}.
Although smart contracts are run and verified in a distributed fashion, their
semantics suggest that one can think of them as of \emph{sequential} programs,
despite the existence of a number of complex interaction patterns including
\eg, reentrancy and recursive calls.
This mental model simplifies both formal and informal reasoning about contracts,
enabling immediate reuse of existing general-purpose frameworks for program
verification~\cite{Filliatre-Paskevich:ESOP13,formal-solidity,Bhargavan-al:PLAS16,ethereum-idris:16}
that can be employed to verify smart contracts written in \eg Solidity~\cite{solidity}
with only minor adjustments.

Although all computations on a blockchain are
deterministic,\footnote{This requirement stems from the way the underlying
Byzantine distributed ledger consensus protocol enables
all involved parties to agree on transaction outcomes.}
a certain amount \emph{non-determinism} still occurs due to
races between transactions themselves (\emph{i.e.} which transactions are
chosen for a given block by the miners).
We will show in that non-determinism can be exploited by adversarial parties
and makes reasoning about contract behavior particularly subtle, reminiscent
to known challenges involved in conventional concurrent programming.

In this paper we outline a model of smart contracts that emphasizes
the properties of their \emph{concurrent} executions.
Such executions can span \emph{multiple} blockchain transactions (within
the same block or in multiple blocks) and thereby violate desired safety
properties that cannot be stated using only the contract's implementation and
local state---precisely what the existing verification methodologies focus
on~\cite{formal-solidity,Bhargavan-al:PLAS16}.  To facilitate the reuse of the
common programming intuition, we propose the following analogy:
\begin{center}
  {\textbf{Accounts using smart contracts in a blockchain\\are like\\
  threads using concurrent objects in shared memory.}}
\end{center}

\paragraph{Threads using concurrent objects in shared memory.~~}
By \emph{concurrent objects} we mean the
broad class of data structures that are employed to exchange data between and
manage the interaction of
multiple \emph{threads} (processes) running concurrently~\cite{Herlihy-Shavit:08}.
Typical examples of concurrent objects are locks, queues, and atomic
counters---typically used via popular libraries such as
\texttt{java.util.concurrent}.  At runtime, these concurrent
objects are allocated in a block of \emph{shared memory} that is accessible
to the running threads.  The behavior resulting from the threads accessing the objects
simultaneously---\emph{i.e.} \emph{interference}---can be extremely
unpredictable and thus extremely difficult to reason about.

Concurrent objects whose implementation does not utilize proper
synchronization (\eg, with \emph{locks} or \emph{barriers}) can
manifest \emph{data races}\footnote{That is, unsynchronized concurrent
  accesses by different threads to a single memory location when at
  least one of those accesses is a write.} under interference leading
to a loss of memory integrity.
Even for race-free objects the observed behavior under interference
may be erroneous from the perspective of one or more clients.  For example,
a particular thread may not ``foresee'' the actions taken by the other
threads with a shared object and thus may not expect for that object to change
in all of the ways that it does change under interference.

\paragraph{Accounts using smart contracts in a blockchain.~~}
Smart contracts are analogous to concurrent objects.  Instead of residing
in a shared memory they live in the blockchain; instead of being used by
threads they are invoked by \emph{accounts} (users or other contracts).  Like
concurrent objects, they have internal mutable state, manage resources (\eg funds),
and can be accessed by multiple parties both within a block and in multiple blocks.
Unlike traditional concurrent objects, a smart contract's methods are atomic
due to the transactional model of computation.  That is, a single call to a
contract (or a chain of calls to a series of contracts calling each other), is
executed \emph{sequentially}---without interrupts---and either terminates
after successfully updating the blockchain or aborts and rolls back to its
previous configuration before the call.
%
%The distributed ledger consensus protocol (implemented differently in
%different cryptocurrencies) will, therefore, ensure that if the
%transaction makes it to the blockchain, it does not violate the
%integrity of state and resources.

The notion of ``atomicity for free'' is deceptive, however, as concurrent
behavior can still be observed \emph{at the level of the blockchain}:
\begin{itemize}

\item The order of the transactions included to a block is not
determined at the moment of a transaction execution, and, thus, the
outcome can largely depend on the ordering with respect to other
transactions~\cite{Luu-al:CCS16}.

\item Several programming tasks require the contract logic to be spread across
several blockchain transactions (\eg, when contracts ``communicate'' with the
world outside of the blockchain), enabling true concurrent behavior.

\item Calling other contracts can be considered
to be a kind of \emph{cooperative multitasking}.  By cooperative multitasking
we mean that multiple threads can run but do not get interrupted unless they
explicitly ``yield''.  That is, a call from contract A to contract B
can be considered to be a yield from contract A's perspective, with contract B
yielding when it returns.  The key point for smart contracts is that
\textbf{contract~B can run code that was unanticipated by contract A's designer},
which makes the situation much closer to a concurrent setting than a typical
sequential one.\footnote{A better term would be ``uncooperative multitasking''
under the circumstances.}
In particular, contract B can modify state that contract A may assume is unchanged during
the call.  This is the essence of The~\dao bug~\cite{dao}, in which contract B
made a call back into contract A to modify A's local state before
returning~\cite{Luu-al:CCS16}.
However, reentrancy is not the only way this kind of error can manifest,
since:
\item It is not difficult to imagine a scenario in which a certain
contract is used as a \emph{service} for other parties (users and
contracts), managing the access to a shared resource and, in some
sense, serving as a concurrent library.  As multi-contract
transactions are becoming more ubiquitous, various interference
patterns can be observed and, thus, should be accounted for.
\end{itemize}

\paragraph{Our goals and motivation.~}

Luckily, the research in concurrent and distributed programming
conducted in the past three decades provides a large body of
theoretical and applied frameworks to code, specify, reason about, and
formally verify concurrent objects and their implementations.
The goal of this paper is thus twofold.
First, we are going to provide a brief overview of some known
concurrency issues that can occur in smart contracts, characterizing the
problems in terms of more traditional concurrency abstractions.
Second, we are aiming to build an intuition for ``good'' and ``bad''
contract behaviors that can be identified and verified/detected
correspondingly, using existing formal methods developed for reasoning
about concurrency.

% \subsection{Paper outline}
% \label{sec:paper-outline}

% \marginpar{We should rework this section once the rest of the paper is finalized.
% Also the abstract.  ;-)}

% In the remainder of the paper, we will demonstrate a number of
% concurrency-related issues exhibited by seemingly correct
% contracts. Specifically, we focus on the \emph{lack of
%   synchronization} (Section~\ref{sec:sync}), \emph{ownership and
%   sharing} (Section~\ref{sec:sep}), \emph{fine-grained operations}
% (Section~\ref{sec:overview}), \emph{composition}, \emph{reentrancy}
% and \emph{liveness} (Section~\ref{sec:discussion}).
% %
% For each aspect, we will refer to existing formal methodologies and
% tools that already address the potential problems in a concurrent
% setting.
% %
% We will then sketch a general specification methodology simplifying
% the formal reasoning about interaction properties of Ethereum-style
% contracts (Section~\ref{sec:formal}) and elaborate on the related
% results from smart contract verification to date
% (Section~\ref{sec:related}), concluding with the perspectives, enabled
% by the observed analogy (Section~\ref{sec:conclusion}).

\section{Deployed Examples of \emph{Concurrentesque} Behavior}
\label{sec:overview}

Here we discuss two contracts that have been deployed on the Ethereum blockchain that each illustrate different aspects of concurrent-type behavior.  The \blockking contract, like many others on the Ethereum blockchain today, implements a simple gambling game~\cite{blockking}.  Although \blockking is not heavily used, we study it because it showcases a potential use of the \oraclize service~\cite{oraclize}, which is a service that allows contracts to communicate with the world outside of the blockchain and thus invites true concurrency.  Since the early adopters of the \oraclize service wrote it as a demonstration of the service and has made its source code freely available, it is likely that many other contracts that wish to use \oraclize will mirror it in their implementations.

%Moreover, if the clients of \oraclize themselves misunderstood the concurrent behavior then we believe others will too.

The second example we discuss is the widely-studied bug in the \dao contract~\cite{daoabout}.  The \dao established an owner-managed venture capital fund with more than 18,000 investors; at its height it attracted more than 14\% of all Ether coins in existence at that time.  The subsequent attack on it cost investors approximately 3.6 million Ether, which at that time was worth approximately USD 50 million.  The \dao employed what we call ``uncooperative multitasking'', in that when the \dao sent money to a recipient then that recipient was able to run code that interfered (via reentrancy) with the \dao's contract state that the \dao assumed would not change during the call.

\subsection{The \blockking contract}
\label{sec:blockking}

The gamble in \blockking works as follows. At any given time there is a designated ``Block King'' (initially the writer of the contract).  When money is sent to the contract by a sender $s$, a random number $j$ is generated between 1 and 9.  If the current block number modulo 10 is equal to $j$ then $s$ becomes the new Block King.  Afterwards, the Block King gets sent a percentage of the money in the contract (from 50\% to 90\% depending on various parameters), and the writer of the contract gets sent the balance.

Generation of good quality random numbers is often difficult in deterministic systems, especially in a context in which all data is publicly stored---and in which there are financial incentives for attackers.  Accordingly, \blockking utilizes the services of a trusted party, Wolfram Alpha, to generate its random numbers using the \oraclize service.  Assuming \oraclize is well-behaved, this strategy for random number selection should be very difficult for attackers to predict.

\begin{figure}[t]
\begin{lstlisting}[language=Solidity,numbers=left,firstnumber=293,mathescape=true,  basicstyle=\scriptsize\ttfamily]
function enter() {
 // 100 finney = .05 ether minimum payment otherwise refund payment and stop contract
 if (msg.value < 50 finney) {
  msg.sender.send(msg.value);
  return;
 }
 warrior = msg.sender; $\label{code:bkentvarsetbegin}$
 warriorGold = msg.value;
 warriorBlock = block.number; $\label{code:bkentvarsetend}$
 bytes32 myid =
  oraclize_query(0,"WolframAlpha","random number between 1 and 9"); $\label{code:bkcalloracle}$
} $\label{code:bkexitenter}$

function __callback(bytes32 myid, string result) { $\label{code:bkcallback}$
 if (msg.sender != oraclize_cbAddress()) throw;
 randomNumber = uint(bytes(result)[0]) - 48; $\label{code:bksetrand}$
 process_payment(); $\label{code:bkcallprocesspayment}$
}

function process_payment() {
\end{lstlisting}
\texttt{...}
\begin{lstlisting}[language=Solidity,numbers=left,firstnumber=339,mathescape=true,basicstyle=\scriptsize\ttfamily]
 if (singleDigitBlock == randomNumber) { $\label{code:bkstartbad}$
  rewardPercent = 50;
  // If the payment was more than .999 ether then increase reward percentage
  if (warriorGold > 999 finney) {
   rewardPercent = 75;
  }	
  king = warrior; $\label{code:bkcrown}$
  kingBlock = warriorBlock;
 } $\label{code:bkendbad}$
\end{lstlisting}
%\texttt{...}
%\begin{lstlisting}[numbers=left,firstnumber=365,mathescape=true,language=Solidity]
%}
%\end{lstlisting}
\caption{\blockking code fragments~\cite{blockking}.}
\label{fig:blockking}
\end{figure}

The code for \blockking is 365 lines long, but the lines of particular interest are given in Figure~\ref{fig:blockking}; line numbers here refer to the actual source code of the contract as given by Etherscan~\cite{blockking}.  The \scode{enter} function is called when money is sent to the contract.  It sets some contract variables (lines~\ref{code:bkentvarsetbegin}--\ref{code:bkentvarsetend}) and then sends a query to the \oraclize service (line~\ref{code:bkcalloracle}).

The \scode{oraclize_query} function raises an event visible in the ``real world'' before returning to its caller, which then exits (line~\ref{code:bkexitenter}).  In the real world the \oraclize servers monitor the event logs, service the request (in this case by contacting the Wolfram Alpha web service), and then make a fresh call into the originating contract at a designated callback point (line~\ref{code:bkcallback} in \blockking).  Between the event and its callback, many things can occur, in the sense that the the blockchain can advance several blocks between the call to \scode{oraclize_query} and the resumption of control at \scode{__callback}.  During this time the state of the blockchain, and even of the \blockking contract itself, can have changed drastically.  In other words, \emph{this is true concurrent behavior on the blockchain}.

What can go wrong?  Suppose that multiple gamblers wish to try their luck in a short period of time (even within the same block).  The contract makes no attempt to track this behavior.  Accordingly, each new contestant will overwrite the previous one's data (the critical \scode{warriorBlock} and \scode{warrior} variables) in lines~\ref{code:bkentvarsetbegin}--\ref{code:bkentvarsetend}.  When the callbacks do eventually occur, the last contestant in the batch %(whose address will be contained in the \scode{warrior} variable)
will enjoy multiple chances to win the throne curtesy of the earlier contestants in that batch who payed for the other callbacks!  The culprit is lines~\ref{code:bkstartbad}--\ref{code:bkendbad} from the \scode{process_payment} function, called as the last line of the \scode{__callback} function in line~\ref{code:bkcallprocesspayment}.

Each time the \scode{process_payment} function is called the least significant digit of \scode{warriorBlock} is computed and stored into the variable \scode{singleDigitBlock}.\footnote{For reasons that seem rather strange to us, this modulus is computed very inefficiently in lines 315--338 of the contract, which we elide to save space.}  Each time the \scode{process_payment} function is called by \scode{__callback} he has a new chance to match the random number in line~\ref{code:bkstartbad}.  If the numbers do match, then that final contestant is crowned on line~\ref{code:bkcrown}.

\subsection{The \dao contract}
\label{sec:daocontract}

\begin{figure}[t]
\begin{lstlisting}[numbers=left,firstnumber=1010,mathescape=true,language=Solidity]
 // Burn DAO Tokens
 Transfer(msg.sender, 0, balances[msg.sender]);
 withdrawRewardFor(msg.sender); // be nice, and get his rewards $\label{code:daowithdraw}$
 totalSupply -= balances[msg.sender];
 balances[msg.sender] = 0; $\label{code:daozerobalance}$
 paidOut[msg.sender] = 0;
 return true;
}
\end{lstlisting}
\caption{\dao code fragment~\cite{daocontract}.}
\label{fig:daocode}
\end{figure}

The source code for the \dao is 1,239 lines and markedly more complex than \blockking~\cite{daocontract}.  Since much has already been written about this
bug (\eg \cite{dao,Luu-al:CCS16}), we present in Figure~\ref{fig:daocode} only the key lines.
The problem is the order of line~\ref{code:daowithdraw}, which (via a series of 
further function calls) sends Ether to \scode{msg.sender}, and line~\ref{code:daozerobalance},
which zeros out the balance of \scode{msg.sender}'s account.

In a sequential program, reordering two independent operations has no effect on the ultimate behavior
of the program.  However, in a concurrent program the effect of a sequentially-harmless reorder can have significant effect since the order in which operations occur can affect how the threads interfere.
In the \dao, sending the Ether in line~\ref{code:daowithdraw} ``yields'' control,
in some multitasking sense, to any arbitrary (and thus potentially malicious) contract located
at \scode{msg.sender}.

Unfortunately, the \dao internal state still indicates that the account
is funded since its account balance has not yet been zeroed out in line~\ref{code:daozerobalance}.
Accordingly, a malicious \scode{msg.sender} can initiate a second withdrawal by calling back into the
\dao contract, which will in turn send a second payment when control reaches line~\ref{code:daowithdraw}
again.  In fact, the malicious \scode{msg.sender} can then initiate a third, fourth, \etc.  withdrawal, 
all of which will result in payment.  Only at the end is his account zeroed out, after being paid many
multiples of its original balance.

Previous analyses of this bug have indicated that the problem is due to recursion or unintended reentrancy.  In a narrow sense this is true, but in a wider sense what is going on is that sequential code is running in what is in many senses a concurrent environment.

%Contracts/problems to build upon:
%
%\begin{itemize}
%\item The DAO with recovery
%\item \oraclize contract and interplay
%\item Puzzle contract and granularity
%\item the ABA problem an analogues
%\item Linearizability with Ownership Transfer
%\end{itemize}
%
%\todo{Pick one to showcase and the explain the rest in the next section.}

\section{Interference and Synchronization}
\label{sec:sync}

Having showed that concurrent-type behavior exists and causes problems in real 
contracts on the Blockchain, we will now examine other ways that 
our \emph{concurrent-objects-as-contracts} viewpoint can help us understand
how contracts can behave on the blockchain.

%we start from a simple artificial example recast in
%both worlds: shared memory and blockchain.

\subsection{Atomic updates in shared-memory concurrency}

Figure~\ref{fig:counter} depicts a canonical example (presented in a
Java 8-like pseudocode) of a wrongly used concurrent object, which is
supposed to implement an ``atomic'' counter with methods \texttt{get}
and \texttt{set}. The implementation of the concurrent counter on the
left is obviously \emph{thread-safe} (\ie, \emph{data race}-free),
thanks to the use of \jcode{synchronized} primitives~\cite{Goets-al:JUC}.
What is problematic, though, is how an instance of the \jcode{Counter}
class is used in the multithreaded client  code on the right.

Specifically, with two threads running in parallel and their
operations interleaving, the call to \jcode{incr()} within
\jcode{thread2}'s body could happen, for instance, between the
assignment to \jcode{a} and the call \jcode{c.set(a + 1)} within the
\jcode{incr()} call of \jcode{thread1}. This would invalidate the
condition in the following \jcode{assert} statement, making the
overall program fail \emph{non-deterministically} for a certain
execution!

The issue arises because the implementation of \jcode{incr()} on top
of \jcode{Counter} does not provide the \emph{atomicity guarantees},
expected by the client code. Specifically, the code on the right is
implemented in the assumption that there will be \emph{no
  interference} between the statements of \jcode{incr()}, hence the
counter \jcode{c} is going to be incremented by \jcode{1}, and
\jcode{a} and \jcode{b} will be the same by the end of its execution.
Indeed, this is not always the case in the presence of concurrently
running \jcode{thread2}, and not only \jcode{a} and \jcode{b} will be
different, the later call to \jcode{c.set()} will also ``overwrite''
the result of the earlier one.

\begin{figure}[t]
\centering
\begin{tabular}{c@{\ \ \ }|@{\ \ \ }c}
\begin{lstlisting}[language=Java]
class Counter {
  private int x = 0;

  /** Return current value */
  synchronized int get() {
    return x;
  }

  /** Set x to be v */
  synchronized int set(int v) {
    int t = x;
    x = v;
    return t;
  }
}
\end{lstlisting}
&
\begin{lstlisting}[language=Java]
final Counter c = new Counter();

void incr() {
  int a = c.get();
  int b = c.set(a + 1);
  assert (a == b);
}

// In the main method
Runnable thread1 = () ->
  { incr(); }

Runnable thread2 = () ->
  { incr(); }

thread1.run(); thread2.run();
\end{lstlisting}
\end{tabular}
\caption{A concurrent counter (left) and its two-thread client
  application (right).}
\label{fig:counter}
\end{figure}

A better designed implementation of \jcode{Counter} could have instead
provided an \emph{atomic} implementation of \jcode{incr()},
implemented via a version of \emph{fetch-and-increment}
operation~\cite[\S~5.6]{Herlihy-Shavit:08}, via explicit locking, or
by means of Java's \jcode{synchronized} keyword.
However, given the only two methods, \jcode{get} and \jcode{set}, the
implementation of \jcode{Counter} has synchronization properties of an
atomic register whose \emph{consensus
  number}~\cite[\S~5.1]{Herlihy-Shavit:08} (\ie, the number of
concurrent threads that can unambiguously agree on the outcomes of
\jcode{get} and \jcode{set}) is exactly 1. Therefore, it is
fundamentally impossible to implement an atomic incrementation of
\jcode{c} by using only \jcode{get} and \jcode{set}, and without
relying on some additional synchronization, by giving priorities to
certain preordained threads.

Perhaps a bit surprisingly, even though the implementation of
\jcode{Counter} from Figure~\ref{fig:counter} is not flawed by itself,
its weak atomicity properties render it quite useless in the presence
of an unbounded number of threads, making it virtually impossible to
make any \emph{stable} (\ie, resilient with respect to concurrent
changes) assumptions about its internal state.

\subsection{Atomic updates in concurrent blockchain transactions}

\begin{figure}[t]
\centering
\begin{tabular}{c@{\ \ }|@{\ \ }c}
\begin{lstlisting}[language=Solidity]
contract Counter {
  address public id;
  uint private balance;

  function get() returns (uint) {
    return balance;
  }

  function set() returns (uint) {
    uint t = balance;
    balance = msg.value;
    msg.sender.send(t);
    return t;
  }}
\end{lstlisting}
&
\begin{lstlisting}[language=Solidity]
// ...
// Same code as in Counter

function testAndSet(uint expected)
  returns (uint) {
  uint t = balance;
  if (t == expected) {
    balance = msg.value;
    msg.sender.send(t);
    return t;
  } else {
    throw;
  }
}
\end{lstlisting}
\end{tabular}
\caption{A counter contract (left) and a synchronizing
  \scode{testAndSet} method (right).}
\label{fig:scounter}
\end{figure}

The left part of Figure~\ref{fig:scounter} shows a smart contract,
implemented in Solidity~\cite{solidity}, with functionality and methods
reminiscent to those of an atomic concurrent counter. The function
\scode{get} allows one to query the contract for the current balance,
associated with some fixed address \scode{id}, whereas the \scode{set}
function allows one to update balance with the new balance, taken from
the message via \scode{msg.value}, sending back the old amount and
returning it as a result.

Since the bodies of both \scode{get} and \scode{set} are going to be
executed sequentially in the course of some transactions, neither
there is any need to synchronized them, nor there is any explicit way
to do so in Solidity.
However, it is not difficult to observe that as an implementation of
the simplest possible storage (\eg, for some \scode{id}-related
funds), used by multiple different parties to update it's balance, the
\scode{Counter} contract is as useless as its Java counterpart from
Figure~\ref{fig:counter}.

For instance, imagine that two parties, unaware of each other try to
increment the amount, stored by an instance of \scode{Counter} by a
certain value. Since the contract does not provide a way for them to
do it in one operation, they will have to first query the amount via
\scode{get} and then try to change it via \scode{set} function,
following the same pattern as the implementation of \jcode{incr} from
Figure~\ref{fig:counter}. Indeed, both these calls can be accomplished
in a single transaction, which would make the execution
sequential. However, because of the limited gas
requirement,\footnote{This is a standard way in Ethereum to ensure
  that execution of a contract terminates: by supplying it with a
  limited amount of ``gas'', used as a fuel for execution steps.} it
is ill-advised to call more than one external contract in the course
of execution. Furthermore, the call to \scode{get} can be performed by
a client, external to the blockchain, which would mean that the
consecutive calls to \scode{get} and \scode{set} will end up in
\emph{two different} transactions.
If this is the case, those calls might interfere with other
transactions, launched by multiple parties trying to modify
\scode{Counter} at the same time, making us face the familiar problem:
the result of calling the function \scode{set} cannot be predicted out
of the local observations.

The cause of the described problem, both in the shared-memory and
blockchain cases, is the lack of \emph{strong synchronization
  primitives}, allowing one to simultaneously observe and manipulate
with the counter in the presence of concurrent executions.
One solution to the problem, which would make it possible to increment
the counter atomically, is to enhance the counter with the
\scode{testAndSet} function (right part of Figure~\ref{fig:scounter}).
This function implements the check/update logic similar to the
\emph{compare-and-swap} primitive~\cite[\S~5.8]{Herlihy-Shavit:08},
(known as \jcode{CMPXCHG}, on the Intel x86 and Itanium
architectures), as a way to implement synchronization between multiple
threads. The consensus number of \scode{testAndSet} (and some other
similar \emph{Read-Modify-Write} primitives) is known to be $\infty$,
hence it is strong enough to allow an arbitrary number of concurrent
parties agree on the outcome of the operation.

\paragraph{Notes on formal reasoning and verification.}

The modern formal approaches for runtime concurrency verification,
based on exploring dynamic execution traces and summarizing their
properties, provide efficient tools for detecting the violations of
atomicity assumptions, and the lack of
synchronization~\cite{Lin-Dig:ICST13}.
For instance, by translating our contract to the corresponding
shared-memory concurrent object, one would be able to use the existing
tools to summarize its traces~\cite{Emmi-al:POPL16}, thus, making it
possible to observe undesired interaction patterns.

% This is not the only way to control interference at a shared
% object. An alternative solution would be to enhance the counter with a
% number of additional fields, limiting manipulations with it to a fixed
% set of threads, or just one thread. We elaborate on this approach in
% the next section.

% Alternatively, one could implement a version of a lock, as a part of
% the counter or an additional contract, which would be used as a
% semaphore for accessing \scode{Counter}'s instance, synchronizing the
% concurrent accesses. Engineering such a lock is feasible, but would
% require the counter to be modified, in order to avoid accesses that to
% modify its state without acquiring the lock---a challenge discussed in
% Section~\ref{sec:further-examples}.

% \begin{comment}

% \begin{enumerate}

% \item What are other potential pitfalls in concurrency and how it is
%   manifested in the contracts:

%   \begin{itemize}
%   \item Breaking observational refinement (DAO with recovery, King of
%     ether);
%   \item Non-atomicity (Oraclize);
%   \item ABA problem (lost updates);
%   \item Lack of proper specification;
%   \item Reasoning about clients and horizontal composition;
%   \end{itemize}

% \item given the gas semantics, even simple contracts should be
%   reconsidered in this setting;

% \item Good news: solutions and tools;

% \item Related problems: reasoning about loops;

% \item Initial development a STS model of smart contracts;
% \end{enumerate}

% \end{comment}

\section{State Ownership and Permission Accounting}
\label{sec:sep}

A different way to prohibit the unwelcome interference on a contract's
state is to engineer a tailored permission accounting discipline,
controlling the set of operations allowed for different parties.

Let us first notice that the problems exhibited by the two-thread
example in Figure~\ref{fig:counter} and preventing one from asserting
anything about its state \jcode{x} could be avoided if we enforced a
restricted access discipline: for instance, by stating that at any
moment at most one thread can query/modify its state. This would grant
the corresponding thread an exclusive
\emph{ownership}~\cite{OHearn:TCS07} over the object, thus, justifying
any assertions made locally from this thread about the object's state.

\begin{figure}[t]
\centering
\begin{tabular}{c@{\ \ \ }|@{\ \ \ }c}
\begin{lstlisting}[language=Solidity,basicstyle=\scriptsize\ttfamily]
contract Counter {
  address public owner;
  uint private balance;

  modifier byOwner() {
    if (msg.sender != owner) throw;
    _
  }

  function get() external byOwner 
    returns (uint) {
    return balance;
  }

  function set() external byOwner  
    returns (uint) {
    uint t = balance;
    balance = msg.value;
    msg.sender.send(t);
    return t; 
  }}
\end{lstlisting} 
&
\begin{lstlisting}[language=Solidity,basicstyle=\scriptsize\ttfamily]
// Same declarations as in Counter

mapping (address => bool) readers;

// Initialized with 0x0
address writer;

modifier canRead() {
  if (msg.sender != writer ||
      !readers[msg.sender]) throw;
   _
}

modifier canWrite() {
  if (msg.sender != writer) throw;
  _
}

function acquireReadLock() returns (bool) {
  if (writer == 0x0) {
    readers[msg.sender] = true; 
  } else return false;
}

// ... Other synchronization primitives

\end{lstlisting} 
\end{tabular}
\caption{An exclusively-owned (left) and Read/Write-locked (right) contract.} 
\label{fig:ocounter}
\end{figure}

The unique ownership is traditionally ensured in Ethereum's contracts
by disallowing any other party, but a dedicated \emph{owner}, make
critical changes in the contract state. For instance,
Figure~\ref{fig:ocounter} (left) shows an altered version of the
\scode{Counter} contract, so no other party can interact with it but
its ``owner''.
The ownership discipline is enforced by Solidity's mechanism of
\scode{modifier}s, allowing one to provide custom dynamically checked
pre-/postconditions for functions. In our example, the \scode{byOwner}
modifier will enforce that the functions \scode{get} and \scode{set}
will be only invoked on behalf of a fixed party---the \scode{owner} of
the contract.

This is a rather crude solution to the interference problem, as it
would mean to exclude any concurrent interaction at a contract
whatsoever. It is quite illuminating, though, from a perspective on
thinking of contracts as concurrent objects, allowing us to
immediately apply our analogy: \emph{{accounts are threads}}. Indeed,
by imposing a specific ownership discipline on a contract as shown in
Figure~\ref{fig:ocounter} is similar to enhancing its Java counterpart
with an explicit check of \jcode{Thread.currentThread().getId()}.

Let us now try to push the analogy between accounts and threads a bit
further by designing a version of a counter with more elaborated
access rights. 
In particular, we are going to ensure that as long as there are
accounts (aka ``threads'') ``interested'' in having its value
immutable (as their internal logic might rely on its immutability), no
other party may be allowed to modify it. Similarly, if at the moment
there is exactly one party that holds a unique permission to modify
the counter, no other parties may be allowed to read it. The solution
to this synchronization problem is well-known in a concurrency
community by the name \emph{Read/Write
  lock}~\cite{Bornat-al:POPL05}. Its implementation requires keeping
track of threads currently reading and writing to the shared object,
so a thread should explicitly \emph{acquire} the corresponding
permission before performing a read/write operation, and then should
\emph{release} it upon finishing.  

The right part of Figure~\ref{fig:ocounter} shows the essential
fragments of the Read/Write-locked contract implementation. The two
new fields, \scode{readers} and \scode{writer} keep track of the
currently active readers and writers. The new modifiers
\scode{canRead} and \scode{canWrite} are to be used for the omitted
\scode{get} and \scode{set} operations correspondingly. Finally,
\scode{acquireReadLock} allows its caller to acquire the lock as long
as there is no active writer in the system, by registering it in the
\scode{readers} mapping.

As we can see, the accounts-as-threads is a rather powerful analogy,
suggesting a number of solutions to possible synchronization problems
that can be taken verbatime from the concurrency literature.
The only drawback of the presented solution is the fact that it is
rather monolithic: the contract now combines the functionality of the
data structure (\ie, the counter) and that of a synchronization
primitive (\ie, a lock). We will discuss possible ways to improve the
modularity of the implementation in Section~\ref{sec:discussion}. 

\paragraph{Notes on formal reasoning and verification.~~}

Formal reasoning about permission accounting and separation of state
access is a long studied topic in the shared-memory concurrency
literature (see, \eg,~\cite{Brookes-OHearn:SIGLOG16} for an overview).
Formalisms, such as Concurrent Separation Logic
and~\cite{OHearn:TCS07} Fractional/Counting
permissions~\cite{Bornat-al:POPL05} provide a flexible way to define
the abstract ownership discipline and verify that a particular
implementation follows it faithfully. For instance, our Read/Write
lock contract can be formally proven \emph{safe} (\ie, prohibiting
concurrent write-modifications) using a formal model of permissions by
Bornat \etal~\cite{Bornat-al:POPL05}.

\section{Discussion}
\label{sec:discussion}

% We have considered several major behavioral aspects implied by the
% concurrent nature of smart contracts in a blockchain: preemptive
% interaction, atomicity, and permission accounting.
% %
% We will now discuss two related correctness concerns, which might
% benefit from the concurrency perspective.

\subsection{Composing the contracts}
\label{sec:composing-contracts}

The locking contract ``pattern'', considered in Section~\ref{sec:sep},
has a significant drawback: its design is \emph{non-modular}. That is,
the locking machinery is implemented by the contract itself rather
than by a third-party library. This is at odds with good practices of
software engineering, in which it is advised to implement
synchronization primitives, such as ordinary and reentrant locks, as
standalone libraries, which can be used for managing access
client-specific resources.

But once the lock logic is factored out of the contract, the reasoning
about the contract's behavior becomes significantly more difficult,
as, in order to prove the preservation of its internal invariants, one
needs to be aware of the properties of the extracted locking protocol,
such as, \eg, uniqueness of a writer, which are external to the
contract. In other words, verification of a contract can no longer be
conduced in an \emph{isolated} manner and will require building a
model that allows reasoning about a contract interacting with other,
rigorously specified contracts.
The idea of disentangling the logic of contracts is not inherent to
our concurrent view and is paramount in the existing good practices of
contract development. For instance, the same idea is advocated as a
way to implement \emph{upgradable} contracts in Ethereum through
introducing and additional level of
indirection~\cite{upgradable-contracts}. Having a ``contract
factory'', implemented as another contract, which can be invoked by
any party, poses verification challenges similar to those of proving
the safety properties of \emph{higher-order} concurrent object (\ie,
an object, that is manipulating with other
objects)~\cite{Hendler-al:SPAA10}.

The idea of compositional reasoning and verification of
mutually-dependent and higher-order concurrent objects using
concurrency logics has been a subject of a large research body in the
past
decade~\cite{Turon-al:ICFP13,Sergey-al:PLDI15,DinsdaleYoung-al:ECOOP10,Svendsen-al:ESOP13}. Most
of those approaches focus on a notion of \emph{protocol}, serving as
an abstract interface of an object's behavior in the presence of
concurrent updates, while hiding low-level implementation details
(\ie, the actual code). We believe, that by leveraging our analogy, we
will be able to develop a method for modular verification of such
multi-contract interactions.

\subsection{Liveness properties}
\label{sec:other-conc-issu}

With the introduction of locks and exclusive access, another
concurrency-related issue arises: reasoning about \emph{progress} and
\emph{liveness} properties of contract implementations.
For instance, it is not difficult to imagine a situation, in which a
particular account, registered as a ``reader'' in our example from
Figure~\ref{fig:ocounter}, might never release the reader-lock, thus,
blocking everyone else from being able to change the contract's state
in the future. The liveness in this setting would mean that
\emph{eventually something good happens}, meaning that any party is
properly incentivised to release the lock. In a concurrency
vocabulary, such an assumption can be rephrased as \emph{fairness} of
the system scheduler, making it possible to reuse existing proof
methods for modular reasoning about progress~\cite{Liang-Feng:POPL16}
and termination~\cite{Gotsman-al:POPL09} in of single- and
multi-contract executions.

\section{Related Work}
\label{sec:related}

Formal reasoning about smart contracts is an emerging and exciting
topic, and suitable abstractions for describing a contract's behavior
are a subject of active research.
In this section, we relate our observations to the existing results in
formalizing and verifying contract properties, outlining promising
areas that would benefit from our concurrency analogy.

\subsection{Verifying contract implementations}

Since the \dao bug~\cite{dao}, the Ethereum community has been focusing
on preventing similar errors, with the aid of general-purpose tools
for program verification. 

At the moment, contracts written in Solidity can be annotated with
Hoare-style pre/postconditions and translated down to OCaml
code~\cite{formal-solidity}, so they become amenable to verification
using the Why3 tool, which uses automation to discharge the generated
verification conditions~\cite{Filliatre-Paskevich:ESOP13}.
This approach is efficient for verifying basic safety properties of
Solidity programs, such as particular variables always being within
certain array index boundaries, and preservation of general contract
invariants (typically stated in a form if linear equations over values
of \scode{uint}-valued variables) at the method boundaries and before
performing external contract calls---precisely what was violated by
the \dao contract.

Bhargavan~\etal have recently implemented a translation from a subset
of Solidity (without loops and recursion)~\cite{Bhargavan-al:PLAS16}
into \fstar---a programming language and verification framework, based
on dependent types~\cite{Swamy-al:ICFP11}. They also provided a
translator from EVM bytecode to \fstar programs. Both these approaches
made it possible to use \fstar as a uniform tool for verification of
contract properties, such as invariant preservation and absence of
unhandled exceptions, which were encoded as an effect via \fstar's
support for indexed Hoare monad~\cite{Swamy-al:POPL16}.
A similar approach to specify the behavior of contracts and based on
dependent types has been adopted by Pettersson and
Edstr\"{o}m~\cite{ethereum-idris:16}, who implemented a small
effect-based contract DSL as a shallow embedding into
Idris~\cite{Brady:ICFP13}, with the executable code extracted to
Serpent~\cite{serpent}, a Python-style contract language.

Hirai has recently formalized the entire specification of Ethereum
Virtual Machine~\cite{evmverif} in Lem~\cite{Mulligan-al:ICFP14} with
extraction to the Isabelle/HOL proof assistant, allowing mechanized
verification of contracts, compiled to EVM bytecode, for a number of
safety properties, including assertions on mutable state and the
absence of potential reentrancy.
Unlike the previous approaches, Hirai's formalization does not provide
a syntactic way to construct and compose proofs (\eg, via a
Hoare-style program logics), and all reasoning about contract behavior
is conducted out of the low-level execution
semantics~\cite{ethereum-yellow-paper}.

In contrast with these lines of work, which focus predominantly on
\emph{low-level} safety properties and invariant preservation, our
observations hint a more high-level formalism for capturing the
properties of a contract behavior and its communication patterns with
the outside world. In particular, we consider communicating
state-transition systems (STSs)~\cite{Nanevski-al:ESOP14} with
abstract state as a suitable formalism for proving, \eg, trace and
liveness properties of contract executions using a toolset of
established tools, such as TLA+~\cite{Lamport:TLA}. In order to
connect such an abstract representation with low-level contract code,
one will have to prove a
\emph{refinement}~\cite{Abadi-Lamport:LICS88}
between the high-level and the low-level representations, \ie, between
an STS and the code.
In some sense, finding a suitable contract invariant and proving it
via Why3 or \fstar may be considered as proving a refinement between a
\emph{one-state} transition system, such that the only state is what
is described by the invariant, and an implementation that preserves
it.
However, we expect more complicate STSs will be required in order to
reason about contracts with preemptive concurrency.

\subsection{Reasoning about global contract properties}

The observation about some contracts being prone to unintentional or
adversarial misuse due to the interference phenomenon has been made by
Luu~\etal~\cite{Luu-al:CCS16}. They characterised the problem similar
to what's exhibited by our counter example in Section~\ref{sec:sync}
as \emph{transaction-ordering dependency} (TOD), which under our
concurrency analogy can be generalized as a problem of unrestricted
interference. The solution to the TOD-problem, suggested by Luu~\etal,
required changing the semantics of Ethereum transactions, providing a
primitive, similar to our \scode{testAndSet} from
Figure~\ref{fig:scounter}. While the advantage of such an approach is
the absence of the need to modify the already deployed contracts (only the
client code interacting with them needs to be changed), it requires
all involved users to upgrade their client-side applications, in order
to account for the changes.
In essence, Luu~\etal's solution targets a very specific concurrency
pattern: strengthening synchronization, provided by atomic registers,
by adding a blockchain-supported \emph{read-modify-write}
primitive. Realizing the nature of the problem, hinted by our analogy,
might instead suggest alternative \emph{contract-based} solutions, such
as, \eg, engineering a locking proxy contract. The disadvantage of
this approach is, however, the need to foresee this behavior at the
moment of designing and deploying a contract. That said, such an
ability to model this behavior is precisely what, we believe, our
analogy enables.

% \todo{Say what's a problem with solidity}

% \begin{enumerate}
% \item Synereo
% \item Bamboo
% \end{enumerate}

% \is{Mention the work by \cite{Luu-al:CCS16} in relation to our example
%   in Section~\ref{sec:sync} and their ``heavy-weight'' solution in
%   contrast with our ``lightweight'' one.}

% \paragraph{Formal reasoning about concurrency.}

% \is{Cite stuff on atomicity checking, referred to by Doug Lea.}

% \paragraph{More properties}

% \is{TODO: how to prove the liveness}

\section{Conclusion}
\label{sec:conclusion}

We believe that our analogy between \emph{smart contracts} and
\emph{concurrent objects} can provide new perspectives, stimulate
research, and allow effective reuse of existing results, tools, and
insights for understanding, debugging, and verifying complex
contract behaviors in a distributed ledger.
As any analogy, ours should not be taken verbatim: on the one hand,
there are indeed issues in concurrency, which seem to be hardly
observable in contract programming; on the other hand, smart contract
implementers should also be careful about notions that do not have
direct counterparts in the concurrency realm, such as gas-bounded
executions and management of funds.

To conclude, we leave the reader with several speculations, inspired
by our observations, but neither addressed nor disproved:

\begin{itemize}

\item A common concurrency challenge in non garbage-collected
  languages is to track the uniqueness of heap locations, which can be
  later reclaimed and repurposed---an issue dubbed \emph{the ABA
    problem}~\cite{Dechev-al:ISORC10}. With the lack of due caution,
  the ABA problem may lead to the violation of the object's state
  integrity. Can we imagine a similar scenario in a multi-contract
  setting?

\item Continuing the analogy, if one sees a blockchain as a shared
  state, then the mining protocol defines the priorities for
  scheduling. Can we leverage the insights from efficient concurrent
  thread management in order to analyze and improve the existing
  distributed ledger implementations?

\item \emph{Linearizability}~\cite{Herlihy-Wing:TOPLAS90} (aka
  \emph{atomicity}) is a standard notion of correctness for specifying
  high-level behavior of lock-free concurrent objects. What would be
  an equivalent de-facto notion of consistency for composite contracts
  with multi-transactional operations, such as \blockking?

\end{itemize}

\bibliographystyle{abbrv}
\bibliography{references,proceedings}

\end{document}